\documentclass[journal,twoside,10pt]{IEEEtran}

\usepackage[cmex10]{amsmath}
\usepackage{amssymb}
\usepackage{bm}
\usepackage{mathrsfs}
\usepackage{graphicx}
\usepackage{accents}
\usepackage{epsfig}
\usepackage{subcaption}
\usepackage{paralist}
\usepackage{cite}
\usepackage{setspace}
\usepackage{cancel} 
\usepackage{color}
\usepackage{wrapfig}
\usepackage{xspace}
\usepackage[colorlinks, citecolor=black, linkcolor=black, linktocpage=true]{hyperref}
\usepackage{bookmark}	 
\usepackage{float}
\usepackage{booktabs}
\usepackage{array}
\usepackage{fixltx2e}
\MakeRobust{\underrightarrow}
\usepackage{multirow}

\newcommand{\dt}{\mathrm{d}t} 

\newcommand{\norm}[1]{\left\Vert#1\right\Vert} 
\newcommand{\abs}[1]{\left\vert#1\right\vert} 
\newcommand{\mc}[1]{\mathcal{#1}}  

\newcommand{\bma}[1]{\left[\begin{array}{#1}}
\newcommand{\ema}{\end{array}\right]}

\DeclareMathAlphabet{\mbf}{OT1}{ptm}{b}{n}
\newcommand{\mbs}[1]{{\boldsymbol{#1}}}
\newcommand{\mbc}[1]{ \boldsymbol{\mathcal{#1}} } 

\newcommand{\mbfbar}[1]{{\bar{\mbf{#1}}}}

\newcommand{\mbftilde}[1]{{\tilde{\mbf{#1}}}}

\def\fdotb{{\raisebox{-0.6ex}{ \kern0.2ex\raisebox{0.8ex}{\tiny $\hspace*{-1ex}\circ$}}}}
\def\fddotb{{\raisebox{-0.6ex}{ \kern0.2ex\raisebox{0.8ex}{\tiny $\hspace*{-1ex}\circ\circ$}}}}

\newcommand{\innerp}[2]{\left\langle #1 , #2 \right\rangle}
\newcommand{\ket}[1]{\left\langle #1 \right\rangle}

\newcommand{\trans}{{\ensuremath{\mathsf{T}}}} 
\newcommand{\trace}{ {\ensuremath{\mathrm{tr}}} } 
\newcommand{\cone}{{\ensuremath{\mathrm{cone}}}}

\newcommand{\beq}{\begin{equation}}
\newcommand{\eeq}{\end{equation}}
\newcommand{\bdis}{\begin{displaymath}}
\newcommand{\edis}{\end{displaymath}}
\newcommand{\beqarray}{\begin{eqnarray}}
\newcommand{\eeqarray}{\end{eqnarray}}
\newcommand{\beqarraynn}{\begin{eqnarray*}}
\newcommand{\eeqarraynn}{\end{eqnarray*}}

\newenvironment{packed_enum}{
\begin{enumerate}
  \setlength{\itemsep}{1pt}
  \setlength{\parskip}{0pt}
  \setlength{\parsep}{0pt}
}{\end{enumerate}}

\newtheorem{theorem}{Theorem}[section] 
\newtheorem{lemma}{Lemma}[section]
\newtheorem{corollary}{Corollary}[section]
\newtheorem{remark}{Remark}
\newtheorem{problem}{Problem}[section]

 \newcommand{\proof}{\begin{IEEEproof}}
 \newcommand{\qed}{\end{IEEEproof}}

\usepackage{adjustbox}
\usepackage{tikz}
\usetikzlibrary{shapes,arrows,calc,intersections,through,fit,
    decorations.markings,shapes.geometric,backgrounds}
\usepackage{tkz-euclide}
\usetkzobj{all}

\tikzstyle{decision} = [diamond, draw, fill=blue!20, 
    text width=4.5em, text badly centered, node distance=3cm, inner sep=0pt]
\tikzstyle{block} = [rectangle, draw, fill=blue!20, 
    text width=7em, text centered, rounded corners, minimum height=4em, node distance=5cm]
\tikzstyle{blockEnd} = [rectangle, draw, fill=red!20, 
    text width=7em, text centered, rounded corners, minimum height=4em, node distance=5cm]
\tikzstyle{blockStart} = [rectangle, draw, fill=green!20, 
    text width=7em, text centered, rounded corners, minimum height=4em, node distance=5cm]
\tikzstyle{line} = [draw, -latex',decoration={markings,mark=at position 1 with
    {\arrow[scale=1.3,>=triangle 45]{>}}},postaction={decorate}]
\tikzstyle{lineDoubleArrow} = [draw, -latex',decoration={markings,mark=at position 0 with
    {\arrowreversed[scale=1.3,>=triangle 45]{>}},mark=at position 1 with
    {\arrow[scale=1.3,>=triangle 45]{>}}},postaction={decorate}]
\tikzstyle{cloud} = [draw, ellipse,fill=red!20, node distance=3cm,
    minimum height=2em]

\tikzstyle{block_bd} = [draw, rectangle, 
    minimum height=2em, minimum width=3em]
\tikzstyle{sum} = [draw, circle, node distance=5mm,,inner sep=3pt]
\tikzstyle{input} = [coordinate]
\tikzstyle{output} = [coordinate]
\tikzstyle{pinstyle} = [pin edge={to-,thin,black}]

\tikzset{
  basic box/.style = {
    shape = rectangle,
    align = center,
    draw  = #1,
    fill  = #1!25,
    rounded corners,
    inner sep=10pt},
  header node/.style = {
    Minimum Width = header nodes,
    font          = \strut\Large\ttfamily,
    text depth    = +0pt,
    fill          = white,
    draw},
  header/.style = {%
    inner ysep = +1.5em,
    append after command = {
      \pgfextra{\let\TikZlastnode\tikzlastnode}
      node [header node] (header-\TikZlastnode) at (\TikZlastnode.north) {#1}
      node [span = (\TikZlastnode)(header-\TikZlastnode)]
        at (fit bounding box) (h-\TikZlastnode) {}
    }
  },
  hv/.style = {to path = {-|(\tikztotarget)\tikztonodes}},
  vh/.style = {to path = {|-(\tikztotarget)\tikztonodes}},
  fat blue line/.style = {ultra thick, blue}
}

\begin{document}

\title{Interior-Conic Polytopic Systems Analysis and Control}

\author{Alex~Walsh, \emph{Student Member, IEEE}, and James~Richard~Forbes, \emph{Member, IEEE}%
\thanks{A.\,Walsh is with the Department of Aerospace Engineering, The University of Michigan, FXB Building, 1320 Beal Street, Ann Arbor, Michigan, 48109-2140 USA.
	}%
\thanks{J.\,R.\, Forbes is with the Department of Mechanical Engineering, McGill University, Macdonald Engineering Building, 817 Sherbrooke Street West, Montr\'{e}al, Qu\'{e}bec, H3A 0C3, Canada. Email: james.richard.forbes@mcgill.ca.
       }
\vspace{-8mm} }

\markboth{DRAFT: December 2017}
{Walsh and Forbes: Interior-Conic Polytopic Systems Analysis and Control}

\maketitle

\begin{abstract}
Linear parameter varying (LPV) analysis and controller synthesis theory rooted in the small gain and passivity framework currently exist. The study of conic systems encompasses both small gain and passivity properties, and herein, analysis and controller synthesis for polytopic conic systems is considered. Linear matrix inequality constraints are given to provide conic bounds for polytopic systems. In addition, given controller conic bounds, a control synthesis method is introduced. The polytopic conic controller is demonstrated on a heat exchanger in simulation, and compared to existing LPV control design techniques.
\end{abstract}


\section{Introduction}
\label{sec:introduction}

Input-output properties and associated stability theorems are indispensable tools to a control systems engineer. 
In particular, passivity and small gain properties, as well as their associated input-output stability theorems, are well-known and heavily used in robust, nonlinear, and optimal control.
%
The Passivity Theorem is commonly used for the robust control of mechanical and electrical systems, especially in the context of robotics~\cite{Ortega:1998qf}. 
Other robust control strategies, including $\mathcal{H}_\infty$ control, make use of the Small Gain Theorem, which has been used to robustly control aerospace~\cite{LavretskyWise2013}, electric, and piezoelectric systems~\cite{Wen2000}.

The Small Gain Theorem is at the heart of robust linear parameter varying (LPV) control. Specifically, the Bounded Real Lemma was extended for application to parameter varying systems, enabling the statement of sufficient conditions to derive stabilizing LPV controllers~\cite[p.~15]{Mohammadpour:2012bk}. Two examples that are particularly relevant to this paper are~\cite{Apkarian:1995fk,Bianchi2011}, where polytopic controller synthesis is discussed. In~\cite{Apkarian:1995fk}, subcontrollers at each vertex are synthesized while simultaneously satisfying a global constraint for stability, whereas in~\cite{Bianchi2011}, subcontrollers at each vertex are first synthesized, and then a global constraint for stability is imposed in an effort to reduce computational complexities.

Initial attempts at deriving LPV controllers had shortcomings such as limits on parameter-variation rates and synthesis complexity, but recent efforts have greatly improved LPV control effectiveness and synthesis tractability~\cite{paper_apkarian_adams_1998,paper_Rugh_Shamma_2000,Mohammadpour:2012bk,Hoffmann:2014qf}. Nonetheless, LPV plant descriptions are often approximations of nonlinear systems, and a controller synthesized to stabilize one of these approximations is not guaranteed to stabilize the original nonlinear system~\cite{Hoffmann:2014qf}. For passive systems, the Passivity Theorem is used to circumvent this problem~\cite{WalshForbes2016b,WalshForbes2016a,Walsh2017vsp}. Strictly passive controllers can guarantee closed-loop input-output stability of passive plants, even if an approximate model is used for controller synthesis.

Despite advances in control theory and controller synthesis rooted in the Passivity and Small Gain Theorems, relying on these two theorems introduces limitations. For instance, relying on the Small Gain Theorem may lead to conservative results owing to the fact that the analysis relies on the supremum of the operator in question. Weighting transfer functions can be used to emphasize certain frequency bands, but the conservatism still remains. When stabilizing systems using the Passivity Theorem, plants that are nominally considered passive can have so-called passivity violations due to discretization, time-delay, and sensor noise~\cite{Forbes2012Synth,Bridgeman2014}. With any one or a combination of these violations, results obtained using the Passivity Theorem may be void. As it turns out, the Conic Sector Theorem is a more general input-output stability theorem, of which small gain and passivity are special cases~\cite{Zames1966}.
Synthesizing a controller with the Conic Sector Theorem can help avoid the conservative nature of controllers designed using the Small Gain Theorem, and provide a remedy for situations where the plant features a passivity violation and the Passivity Theorem is not applicable.

The design of conic controllers has been studied for linear systems~\cite{Joshi2002,Bridgeman2014,bridgeman2014a}. The design approach considered in~\cite{Joshi2002,Bridgeman2014,bridgeman2014a} begins by determining the conic bounds of the plant using the Conic Sector Lemma~\cite{Gupta1994}, then choosing appropriate conic bounds of the controller using the Conic Sector Theorem~\cite{Zames1966}, and finally synthesizing a linear controller that satisfies these conic bounds. A similar approach is undertaken in this paper specific to polytopic systems, which are a type of LPV system. 
One paper investigates the use of conic-sector-based control for LPV systems where conicity is violated for certain values of the parameters~\cite{Sivaranjani:2018aa}. 
The challenges to overcome include the fact that currently there are no LMI conditions to determine conic bounds for polytopic systems, and no synthesis methods for conic polytopic controllers.

In short, the two main contributions of this paper are providing a means to assess the conic bounds of polytopic systems using an LMI as stated in Theorem~\ref{thm:conicPolytopic}, and a method to design polytopic conic controllers in Section~\ref{sec:designConicCtrl}. A third contribution is the demonstration of the conic polytopic controller on a heat exchanger.
Notation and preliminaries are in Section~\ref{sec:prelims}, the derivation of conic bounds for polytopic systems are in Section~\ref{sec:conicPolytopic}, the numerical example is in Section~\ref{sec:numericalExample}, and closing remarks are in Section~\ref{sec:conclusion}.

\section{Notation and Preliminaries}
\label{sec:prelims}

\subsection{Notation}

A function $\mbf{u} \in \mc{L}_2$ if $\left\| \mbf{u} \right\|_2 = \sqrt{ \ket{\mbf{u} , \mbf{u}} } = \sqrt{ \int_0^\infty \mbf{u}^\trans(t) \mbf{u}(t)  dt } < \infty$ and $\mbf{u} \in \mc{L}_{2e}$ if $\left\| \mbf{u} \right\|_{2T} = \sqrt{ \ket{\mbf{u} , \mbf{u}}_T }= \sqrt{ \int_0^T \mbf{u}^\trans(t) \mbf{u}(t) dt } < \infty$, $\forall T \in \mathbb{R}_{\geq 0}$~\cite{book_desoer_vidyasagar}.
The matrix $\mbf{1}$ is the identity matrix. A ``$\star$'' indicates symmetry for off-diagonal terms in a matrix.

\subsection{Conic Systems}
\label{sec:conic}

Consider a square system $\mbc{G}: \mc{L}_{2e} \to  \mc{L}_{2e}$, satisfying
\beq
	-\norm{\mbc{G}\mbf{u}}_{2T}^2 + (a + b) \innerp{\mbc{G}\mbf{u}}{\mbf{u}}_T
		-ab \norm{\mbf{u}}_{2T}^2 \geq  \beta,
		\label{eq:conic}
\eeq
for all $\mbf{u} \in \mc{L}_{2e}$ and $T \in \mathbb{R}_{\geq 0}$, where $\beta \in \mathbb{R}$ and $a,b \in \mathbb{R}$, $a<b$. 
The system $\mbc{G}$ is interior conic, denoted $\mbc{G} \in \cone[a,b]$, if~\eqref{eq:conic} holds for some $a$ and $b$. The system $\mbc{G}$ is strictly interior conic, denoted $\mbc{G} \in \cone(a,b)$, if~\eqref{eq:conic} holds for bounds $a+\delta$ and $b-\delta$ for some $\delta > 0$. 
If $a = -\gamma$ and $b = \gamma$, then~\eqref{eq:conic} reduces to the finite-gain expression 
\beq
    \norm{\mbc{G}\mbf{u}}_{2T} \leq \gamma \norm{\mbf{u}}_{2T} + \beta.
        \label{eq:GgainLessGamma}
\eeq
If $\mbc{G}$ is linear and $\beta = 0$, then~\eqref{eq:GgainLessGamma} implies $\norm{\mbc{G}}_\infty \leq \gamma$, where
\bdis
    \norm{\mbc{G}}_\infty = \sup_{\mbf{u} \in \mc{L}_{2e} \setminus \{\mbf{0}\} } \frac{\norm{\mbc{G}\mbf{u}}_{2T}}{\norm{\mbf{u}}_{2T}}.
\edis

\begin {figure}[t]
\centering
\begin{adjustbox}{width=0.7\columnwidth}
\centering
\small
\tikzstyle{line} = [draw, -latex',decoration={markings,mark=at position 1 with
    {\arrow[scale=0.7,>=triangle 45]{>}}},postaction={decorate}]
\begin{tikzpicture}[auto, node distance=2cm,>=latex',text height=1.5ex,text depth=.25ex,
    scale=0.95, every node/.style={scale=0.95}]]
    \node [input, name=input] {};
    \path (input.east)+(1.5,0) node [sum] (sum1) {};
    \path (sum1.east)+(1.5,0) node [block_bd] (G1) {$\mbc{G}_1$};
    \path (G1.south)+(0,-1) node [block_bd] (G2) {$\mbc{G}_2$};
    \path (G2.east)+(1.3,0)  node [sum] (sum2) {};
    \path (G1.east)+(1.3,0)  node [input] (joint1) {};
    \path (G2.west)+(-1.5,0) node [input] (joint2) {};

    \path [line] (input) -- node [near start, above] {$\mbf{r}_1$}
        node [pos=0.9, above] {$+$} (sum1);
    \path [line] (sum1) -- node [midway, above] {$\mbf{u}_1$} (G1.west);
    \path [line] (G1.east) -- node [near end, above] {$\mbf{y}_1$} ($(G1.east) + (2.5,0)$);
    \path [line] (joint1) -- node [pos=0.9, left] {$+$} (sum2);
    \path [line] (joint2) -| node [pos=0.95, left] {$-$} (sum1);
    \path [line] (G2.west) -- node [pos=0.8, above] {$\mbf{y}_2$} ($(G2.west) + (-2.5,0)$);
    \path [line] (sum2.west) -- node [midway,above] {$\mbf{u}_2$} (G2.east);
    \path [line] ($(G2.east) + (2.5,0)$) -- node [near start,above] {$\mbf{r}_2$}
        node [pos=0.9,above] {$+$} (sum2.east);
\end{tikzpicture}
\end{adjustbox}
\caption{Input-output system block diagram.}
\vspace{-2mm}
\label{fig:IOsysBD}
\end{figure}
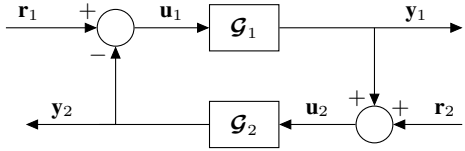

\theorem[Conic Sector Theorem~\cite{Zames1966,Joshi2002}] 
\label{thm:concSector}
Consider the negative feedback interconnection of two square systems, $\mbc{G}_1: \mc{L}_{2e} \to \mc{L}_{2e}$ and $\mbc{G}_2: \mc{L}_{2e} \to \mc{L}_{2e}$, shown in Fig.~\ref{fig:IOsysBD}, where $\mbf{y}_i = \mbc{G}_i \mbf{u}_i$ for $i\in\{1,2\}$. The closed-loop system with inputs $\mbf{r} = [ \mbf{r}_1^\trans \;\; \mbf{r}_2^\trans ]^\trans$ and outputs $\mbf{y} = [ \mbf{y}_1^\trans \;\; \mbf{y}_2^\trans ]^\trans$ is input-output stable if $\mbc{G}_1 \in \cone[a,b]$ for $a<0<b$ and $\mbc{G}_2 \in \cone(-\frac{1}{b},-\frac{1}{a})$. Input-output stability means that if $\mbf{r} \in\mc{L}_2$, then $\mbf{y} \in \mc{L}_2$.

\corollary[Small Gain Theorem~\cite{Brogliato:2007ys}]
\label{thm:smallGain}
Consider the system in Theorem~\ref{thm:concSector}, where $\mbc{G}_1 \in \cone[-\gamma_1,\gamma_1]$ and $\mbc{G}_2 \in \cone[-\gamma_2,\gamma_2]$. If $\gamma_1 \gamma_2 < 1$ and $\mbf{r} \in \mc{L}_{2}$, then $\mbf{u} \in \mc{L}_2$ and $\mbf{y} \in \mc{L}_2$.

\remark Corollary~\ref{thm:smallGain} demonstrates that the Small Gain Theorem is a special case of the Conic Sector Theorem when the systems composing the negative feedback loop are square. However, the Small Gain Theorem is also applicable to non-square systems~\cite{Brogliato:2007ys}.

\section{Conic Polytopic Systems}
\label{sec:conicPolytopic}

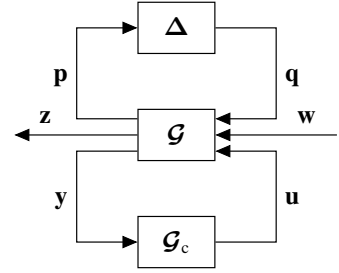
\begin {figure}[t]
\centering
\begin{adjustbox}{width=0.5\columnwidth}
\centering
\small
\tikzstyle{line} = [draw, -latex',decoration={markings,mark=at position 1 with
    {\arrow[scale=0.8,>=triangle 45]{>}}},postaction={decorate}]
\begin{tikzpicture}[auto, node distance=2cm,>=latex',text height=1.5ex,text depth=.25ex,
    scale=0.95, every node/.style={scale=0.95}]]
    \node [input, name=input] {};

    \path (input.east)+(2,0) node [block_bd] (G) {$\mbc{G}$};
    \path (G.south)+(0,-1) node [block_bd] (Gc) {$\mbc{G}_{\textrm{c}}$};
    \path (G.north)+(0,1) node [block_bd] (Delta) {$\mbs{\Delta}$};
    \path (G.east)+(1.5,0) node [input] (output) {};

    \path [line] (G.west) -- node [near end, above] {$\mbf{z}$} (input);
    \path [line] (output) -- node [near start, above] {$\mbf{w}$} (G.east);

    \path [line] ($(G.west) + (0,-0.2)$) -- ++(-0.75,0) |- node [pos=0.25,left] {$\mbf{y}$} (Gc.west);
    \path [line] (Gc.east)-- ++(+0.75,0) |- node [pos=0.25,right] {$\mbf{u}$} ($(G.east) + (0,-0.2)$);

    \path [line] ($(G.west) + (0,+0.2)$) -- ++(-0.75,0) |- node [pos=0.25,left] {$\mbf{p}$} (Delta.west);
    \path [line] (Delta.east)-- ++(0.75,0) |- node [pos=0.25,right] {$\mbf{q}$} ($(G.east) + (0,+0.2)$);
\end{tikzpicture}
\end{adjustbox}
\caption{Standard problem block diagram with an uncertainty block.}
\vspace{-5mm}
\label{fig:sysBlockDiagram}
\end{figure}

\begin {figure}[t]
\begin{subfigure}[First]{0.48\columnwidth}
\centering
\begin{adjustbox}{width=\columnwidth}
\centering
\small
\tikzstyle{line} = [draw, -latex',decoration={markings,mark=at position 1 with
    {\arrow[scale=0.8,>=triangle 45]{>}}},postaction={decorate}]
\begin{tikzpicture}[auto, node distance=2cm,>=latex',text height=1.5ex,text depth=.25ex,
    scale=0.95, every node/.style={scale=0.95}]]
    \node [input, name=input] {};

    \path (input.east)+(2,0.1) node [block_bd] (G) {$\mbc{G}_{\textrm{cl}}$};
    \path (G.north)+(0,1) node [block_bd] (Delta) {$\mbs{\Delta}$};
    \path (G.east)+(1.5,-0.1) node [input] (output) {};

    \path [line] ($(output) + (0,0)$) -- node [near start, above] {$\mbf{w}$} ($(G.east) + (0,-0.1)$);
    \path [line] ($(G.west) + (0,-0.1)$) -- node [near end, above] {$\mbf{z}$} (input);

    \path [line] ($(G.west) + (0,+0.1)$) -- ++(-0.75,0) |- node [pos=0.25,left] {$\mbf{p}$} (Delta.west);
    \path [line] (Delta.east)-- ++(0.75,0) |- node [pos=0.25,right] {$\mbf{q}$} ($(G.east) + (0,+0.1)$);
\end{tikzpicture}
\end{adjustbox}
\caption{Closed-loop control framework relying on the Small Gain Theorem.}
\label{fig:sysBlockDiagramLLFT}
\end{subfigure}
~
\begin{subfigure}[Second]{0.48\columnwidth}
\centering
\begin{adjustbox}{width=\columnwidth}
\centering
\small
\tikzstyle{line} = [draw, -latex',decoration={markings,mark=at position 1 with
    {\arrow[scale=0.8,>=triangle 45]{>}}},postaction={decorate}]
\begin{tikzpicture}[auto, node distance=2cm,>=latex',text height=1.5ex,text depth=.25ex,
    scale=0.95, every node/.style={scale=0.95}]]
    \node [input, name=input] {};

    \path (input.east)+(2,-0.1) node [block_bd] (G) {$\mbc{G}_{\Delta}$};
    \path (G.south)+(0,-1) node [block_bd] (Gc) {$\mbc{G}_{\textrm{c}}$};
    \path (G.east)+(1.5,0.1) node [input] (output) {};

    \path [line] (output) -- node [near start, above] {$\mbf{w}$} ($(G.east) + (0,0.1)$);
    \path [line] ($(G.west) + (0,0.1)$) -- node [near end, above] {$\mbf{z}$} (input);

    \path [line] ($(G.west) + (0,-0.1)$) -- ++(-0.75,0) |- node [pos=0.25,left] {$\mbf{y}$} (Gc.west);
    \path [line] (Gc.east)-- ++(0.75,0) |- node [pos=0.25,right] {$\mbf{u}$} ($(G.east) + (0,-0.1)$);
\end{tikzpicture}
\end{adjustbox}
\caption{Closed-loop control framework relying on the Conic Sector Theorem.}
\label{fig:sysBlockDiagramULFT}
\end{subfigure}
\caption{Upper and lower LFTs for controller synthesis.}
\label{fig:sysBlockDiagramLFT}
\end{figure}
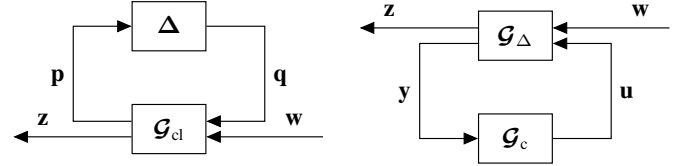

\subsection{System Architecture}

Recall a matrix polytope is defined as the convex hull of a finite number of matrices $\mbf{H}_i$, defined as
\bdis
    \textrm{Co}\{\mbf{H}_i,i=1,\dots,N \} = \left\{ 
        \sum_{i=1}^N s_i \mbf{H}_i \; \Big| \; s_i \geq 0, \; \sum_{i=1}^N s_i = 1 \right\}.
\edis
In this paper, only polytopic systems are examined. Consider the polytopic system shown in Fig.~\ref{fig:sysBlockDiagram}, given by
\begin{align}
	\dot{\mbf{x}} &= \mbf{A} \mbf{x} + \mbf{B}_1 \mbf{w} + \mbf{B}_2 \mbf{u} 
		+ \mbf{B}_3 \mbf{q},
            \label{eq:plant1} \\
	\mbf{z} &= \mbf{C}_1 \mbf{x} + \mbf{D}_{11} \mbf{w} + \mbf{D}_{12} \mbf{u} 
		+ \mbf{D}_{13} \mbf{q}, \\
	\mbf{y} &= \mbf{C}_2 \mbf{x} + \mbf{D}_{21}\mbf{w}, \\
	\mbf{p} &= \mbf{C}_3 \mbf{x} + \mbf{D}_{31} \mbf{w} + \mbf{D}_{32} \mbf{u} 
		+ \mbf{D}_{33} \mbf{q},
        \label{eq:plant4}
\end{align}
where $\mbf{x} \in \mathbb{R}^n$ is the system state, $\mbf{z}\in \mathbb{R}^{n_z}$ is the performance variable, $\mbf{w}\in \mathbb{R}^{n_w}$ is the exogenous signal that can include noise, $\mbf{y} \in \mathbb{R}^m$ is the measurement variable, $\mbf{p} \in \mathbb{R}^{n_p}$ is the input to the uncertainty block, $\mbf{q}\in \mathbb{R}^{n_q}$ is the output of the uncertainty block, and $\mbf{u}\in \mathbb{R}^{m}$ is the system input. Each system matrix is a function of a scheduling signal $s_i$, such that
\beq
\begin{split}
    &\bma{cccc} \mbf{A} & \mbf{B}_1 & \mbf{B}_2 & \mbf{B}_3 \\ 
     \mbf{C}_1 & \mbf{D}_{11} & \mbf{D}_{12} & \mbf{D}_{13} \\
     \mbf{C}_2 & \mbf{D}_{21} & \mbf{0} & \mbf{0} \\
     \mbf{C}_3 & \mbf{D}_{31} & \mbf{D}_{32} & \mbf{D}_{33} \ema \\
    &\qquad = \sum_{i=1}^N s_i(\mbs{\sigma},\mbf{x},t)
    \bma{cccc} \mbf{A}_i & \mbf{B}_{1,i} & \mbf{B}_{2,i} & \mbf{B}_{3,i} \\ 
     \mbf{C}_{1,i}  & \mbf{D}_{11,i} & \mbf{D}_{12,i} & \mbf{D}_{13,i} \\
     \mbf{C}_{2,i} & \mbf{D}_{21,i} & \mbf{0} & \mbf{0} \\
     \mbf{C}_{3,i} & \mbf{D}_{31,i} & \mbf{D}_{32,i} & \mbf{D}_{33,i} \ema,
    \label{eq:polytopicSS}
\end{split}
\eeq
where
\beq
    0  \leq s_i(\mbs{\sigma},\mbf{x},t) \leq 1, \quad
    \sum_{i=1}^N s_i(\mbs{\sigma},\mbf{x},t) = 1.
        \label{eq:ssAsumptions}
\eeq
The scheduling signals $s_i(\mbs{\sigma},\mbf{x},t)$ can be a function of time $t \in \mathbb{R}_{\geq 0}$, a function of state $\mbf{x} \in\mathbb{R}^n$, or a function of an exogenous signal $\mbs{\sigma}\in \mathbb{R}^{n_\sigma}$.

For robust control design using $\mc{H}_\infty$-based techniques, the $\mc{H}_\infty$ norm of the closed-loop system of the plant $\mbc{G}$ and the controller $\mbc{G}_{\textrm{c}}$, denoted $\mbc{G}_{\textrm{cl}}$, is determined. Closed-loop input-output stability with $\mbs{\Delta}$ is guaranteed via Theorem~\ref{thm:smallGain} when $\norm{\mbs{\Delta}}_\infty \norm{\mbc{G}_{\textrm{cl}}}_\infty < 1$. The block diagram for this synthesis is described by Fig.~\ref{fig:sysBlockDiagramLLFT}. Controller synthesis methods using the Small Gain Theorem are extensively covered in the literature. For general reference using LMIs see~\cite{dellerud:2000}, and for polytopic systems, controller synthesis is discussed in~\cite{Apkarian:1995fk}.

Conversely, for robust control design rooted in the conic-systems framework, which is the focus of this paper, the uncertainty and the plant are lumped together as $\mbc{G}_\Delta$, as shown in Fig.~\ref{fig:sysBlockDiagramULFT}. The conic bounds for the system $\mbc{G}_\Delta$ are first determined for $\mbf{u} \mapsto \mbf{y}$, and then appropriate controller bounds are determined via Theorem~\ref{thm:concSector} to ensure closed-loop input-output stability. This is also how robust control based on the passive-systems framework is commonly realized~\cite{Brogliato:2007ys}. The Conic Sector Lemma~\cite{Gupta1994} can be used to determine conic bounds for LTI systems. To apply the Conic Sector Theorem to this robust control design problem the conic bounds of the polytopic system under control must be found. This is addressed in Theorem~\ref{thm:conicPolytopic} in  Section~\ref{sec:conicBounds}.

\subsection{Conic Bounds for Polytopic Systems}
\label{sec:conicBounds}

Conic bounds for a system are defined for a specific input-output pair. For the plant~\eqref{eq:plant1}-\eqref{eq:plant4}, the input-output pair for conic bounds is $\mbf{u} \mapsto \mbf{y}$, with minimal state space realization
\beq
	\mbc{G} :
	\begin{cases}
		&\dot{\mbf{x}} = \mbf{A}(\mbf{s}) \mbf{x} + \mbf{B}_2(\mbf{s}) \mbf{u}\\
		&\mbf{y} =   \mbf{C}_2(\mbf{s}) \mbf{x}
	\end{cases}
		\label{eq:conicPoly}
\eeq
where $\mbf{x} \in \mathbb{R}^{n}$, $\mbf{u} \in \mathbb{R}^m$, $\mbf{y} \in \mathbb{R}^m$, $\mbf{s} = [s_1 \; \cdots \; s_N]^\trans$, and
\beq
	\bma{cc} \mbf{A}(\mbf{s}) & \mbf{B}_2(\mbf{s}) \\  \mbf{C}_2(\mbf{s}) & \mbf{0} \ema
	= \sum_{i=1}^N s_i(\mbs{\sigma},\mbf{x},t)\bma{cc} \mbf{A}_{i} & \mbf{B}_{2,i} \\  \mbf{C}_{2,i} & \mbf{0} \ema,
	\label{eq:polytopicSS}
\eeq
where $\mbf{s}$ satisfies~\eqref{eq:ssAsumptions}. The feedthrough matrix, $\mbf{D}_{22}$, is assumed to be zero as physical systems exhibit some sort of roll-off in gain at higher frequency. If a model does contain a $\mbf{D}_{22}$ matrix, the measurement $\mbf{y}$ can be filtered to remove the feedthrough term.

\lemma
\label{lem:yNorms}
Given $\mbf{y}(t) = \sum_{i=1}^N s_i(\mbs{\sigma},\mbf{x},t) \mbf{y}_i (t)$, where $\mbf{y}(t) \in \mc{L}_{2e}$, $\mbf{y}_i(t) = \mbf{C}_{2,i}\mbf{x}(t)$, and where scheduling signals satisfy~\eqref{eq:ssAsumptions}, then
\beq
	-\norm{\mbf{y}(t)}_{2T}^2 \geq \sum_{i=1}^N -\norm{\sqrt{s_i(\mbs{\sigma},\mbf{x},t)}\mbf{y}_i(t)}_{2T}^2.
		\label{eq:yNormsNeed}
\eeq
\proof See the Appendix. \qed

\begin{theorem}
\label{thm:conicPolytopic}
Consider the system $\mbf{y} = \mbc{G}\mbf{u}$ described by~\eqref{eq:conicPoly}, where $\mbf{A}(\mbf{s})$, $\mbf{B}(\mbf{s})$, and $\mbf{C}(\mbf{s})$ satisfy~\eqref{eq:ssAsumptions} and~\eqref{eq:polytopicSS}.
For $a,b \in \mathbb{R}$, $a<0<b$, if the LMI in $\mbf{P}$
\beq
\begin{split}
    \bma{cc} \mbf{P} \mbf{A}_{i} + \mbf{A}_{i}^\trans \mbf{P} + \mbf{C}_{2,i}^\trans\mbf{C}_{2,i}  & \mbf{P} \mbf{B}_{2,i}- \frac{a+b}{2}\mbf{C}_{2,i}^\trans \\ \star & a b\mbf{1} \ema \\
 = -\bma{c} \mbf{L}_i^\trans \\ \mbf{W}_i^\trans \ema \bma{cc} \mbf{L}_i & \mbf{W}_i \ema \leq 0
\end{split}
\label{eq:gsConicLMI}
\eeq
is satisfied for some real matrices $\mbf{P} = \mbf{P}^\trans > 0$, $\mbf{L}_i$ and $\mbf{W}_i$ for $i=1,\dots,N$, then the system $\mbc{G}$ is in $\cone[a,b]$.

\end{theorem}
\proof
In  this proof, the argument $\mbf{s}$ associated with $(\mbf{A}(\mbf{s}),\mbf{B}(\mbf{s}),\mbf{C}(\mbf{s}))$ is dropped for brevity.
First note that~\eqref{eq:gsConicLMI} implies
\beq
\begin{split}
	\mbf{P} \mbf{A}_{i} + \mbf{A}_{i}^\trans \mbf{P} +\mbf{C}_{2,i}^\trans\mbf{C}_{2,i} &= -\mbf{L}_i^\trans\mbf{L}_i, \\ 
 	\mbf{P} \mbf{B}_{2,i}- \frac{a+b}{2}\mbf{C}_{2,i}^\trans = - \mbf{L}_i^\trans\mbf{W}_i, \qquad &
	a b\mbf{1} = -\mbf{W}_i^\trans\mbf{W}_i.
\end{split}
\label{eq:LMIexpanded}
\eeq
Consider the time derivative of the Lyapunov-like function $\mc{V} = \frac{1}{2} \mbf{x}^\trans \mbf{P} \mbf{x}$,
\beq
	\dot{\mc{V}} = \frac{1}{2}\mbf{x}^\trans(\mbf{P}\mbf{A} + \mbf{A}^\trans \mbf{P}) \mbf{x} +
		\mbf{x}^\trans \mbf{P} \mbf{B} \mbf{u}.
	\label{eq:V_dot}
\eeq
Substituting~\eqref{eq:LMIexpanded} into~\eqref{eq:V_dot} yields
\begin{align}
\dot{\mc{V}} &= \sum_{i=1}^N s_i \bigg[ -\frac{1}{2} \mbf{x}^\trans \mbf{C}_{2,i}^\trans\mbf{C}_{2,i}\mbf{x}
	+  \mbf{x}^\trans\left( \frac{a+b}{2}\right)\mbf{C}_{2,i} \mbf{u}
		\notag \\
	& \quad -\frac{1}{2}\mbf{x}^\trans\mbf{L}_i^\trans \mbf{L}_i\mbf{x}  - \mbf{x}^\trans \mbf{L}_i^\trans \mbf{W}_i \mbf{u}
		\notag  \\
	& \quad -\frac{1}{2} \mbf{u}^\trans\mbf{W}_i^\trans\mbf{W}_i \mbf{u}+ \frac{1}{2} \mbf{u}^\trans\mbf{W}_i^\trans\mbf{W}_i \mbf{u} \bigg] 
		\notag
    \\ 
  &= \sum_{i=1}^N s_i \bigg[ -\frac{1}{2} \mbf{y}_{i}^\trans\mbf{y}_{i}
		  -\frac{1}{2} (\mbf{L}_i\mbf{x} + \mbf{W}_i  \mbf{u})^\trans(\mbf{L}_i\mbf{x} + \mbf{W}_i  \mbf{u})  \bigg]  
		\notag \\
	& \quad   - \frac{1}{2}a b \mbf{u}^\trans \mbf{u}
	+ \left( \frac{a+b}{2}\right) \mbf{y}^\trans \mbf{u}.
		\label{eq:thmLastLine}
\end{align}
Integrating~\eqref{eq:thmLastLine} in time from $0$ to $0<T< \infty$ gives
\begin{align*}
\mc{V}(T) - \mc{V}(0) &= \sum_{i=1}^N \bigg[-\frac{1}{2}\norm{\sqrt{s_i}\mbf{y}_{i}}_{2T}^2 \\
	&\quad-\frac{1}{2} \norm{\sqrt{s_i}\left(\mbf{L}_i\mbf{x} + \mbf{W}_i  \mbf{u} \right)}_{2T}^2 \bigg]  - \frac{a b}{2}\norm{\mbf{u}}_{2T}^2 \\
	&\quad + \left( \frac{a+b}{2}\right)\innerp{\mbf{y}}{\mbf{u}}_{T} .
\end{align*}
Rearranging yields
\beq
\begin{split}
	-\sum_{i=1}^N \norm{\sqrt{s_i}\mbf{y}_{i}}_{2T}^2  + (a + b)\innerp{\mbf{y}}{\mbf{u}}_{T} - ab\norm{\mbf{u}}_{2T}^2 =  \\
	+2(\mc{V}(T) - \mc{V}(0)) +\sum_{i=1}^N \norm{\sqrt{s_i}\left(\mbf{L}_i\mbf{x} + \mbf{W}_i \mbf{u} \right) }_{2T}^2.
\end{split}
\label{eq:thmLineBeforeSimpl}
\eeq
Since $2(\mc{V}(T) - \mc{V}(0)) +\sum_{i=1}^N \norm{\sqrt{s_i}\left(\mbf{L}_i\mbf{x} + \mbf{W}_i \mbf{u} \right) }_{2T}^2 \geq  - 2\mc{V}(0)$, using Lemma~\ref{lem:yNorms} reduces~\eqref{eq:thmLineBeforeSimpl} to
\beq
	-\norm{\mbf{y}}_{2T}^2  + (a + b)\innerp{\mbf{y}}{\mbf{u}}_{T} - ab\norm{\mbf{u}}_{2T}^2 \geq \beta,
		\label{eq:endProofThm}
\eeq
where $\beta = - 2\mc{V}(0)$ only depends on initial conditions. Comparing~\eqref{eq:endProofThm} to~\eqref{eq:conic} proves Theorem~\ref{thm:conicPolytopic}.
\qed

\corollary
\label{col:modifiedConicPoly}
Consider the system $\mbf{y} = \mbc{G}\mbf{u}$ described by~\eqref{eq:conicPoly}.
For $a,b \in \mathbb{R}$, $a<0<b$, if the LMI in $\mbftilde{P}$
\beq
    \bma{cc} \mbftilde{P} \mbf{A}_{i} + \mbf{A}_{i}^\trans \mbftilde{P} +\frac{1}{b}  \mbf{C}_{i}^\trans\mbf{C}_{i}  & \mbftilde{P} \mbf{B}_{i}- \frac{1}{2} \left(\frac{a}{b} + 1\right)\mbf{C}_{i}^\trans \\ \star & a\mbf{1} \ema \\
 \leq 0
    \label{eq:gsConicLMIMod}
\eeq
is satisfied for $i=1,\dots,N$, and $\mbftilde{P} = \mbftilde{P}^\trans > 0$, then the system $\mbc{G}$ is in $\cone[a,b]$.
\proof
Multiplying~\eqref{eq:gsConicLMI} by $\frac{1}{b} > 0$ results in
\bdis
    \bma{cc} \frac{1}{b} \mbf{P} \mbf{A}_{i} + \mbf{A}_{i}^\trans \mbf{P}\frac{1}{b}  + \frac{1}{b} \mbf{C}_{i}^\trans\mbf{C}_{i}  
        & \frac{1}{b}\mbf{P} \mbf{B}_{i}- \frac{1}{2} \left(\frac{a}{b} + 1\right)\mbf{C}_{i}^\trans 
        \\ \star & -a\mbf{1} \ema \leq 0.
\edis
Using the change of variable $\mbftilde{P} = \frac{1}{b} \mbf{P}$ yields~\eqref{eq:gsConicLMIMod}. Since change of variable and multiplication are reversible,~\eqref{eq:gsConicLMIMod} also implies~\eqref{eq:gsConicLMI}. Thus, if $\mbc{G}$ satisfies Corollary~\ref{col:modifiedConicPoly}, then $\mbc{G}$ also satisfies Theorem~\ref{thm:conicPolytopic}.
\qed

Both Theorem~\ref{thm:conicPolytopic} and Corollary~\ref{col:modifiedConicPoly} are sufficient conditions for $\mbc{G} \in \cone[a,b]$. Corollary~\ref{col:modifiedConicPoly} is a reformulation of Theorem~\ref{thm:conicPolytopic}, and may be useful in situations with a large $b$. A large $b$ can cause~\eqref{eq:gsConicLMI} to become ill-conditioned, whereas a large $b$ does not cause problems for~\eqref{eq:gsConicLMIMod}.

\subsection{Determining Conic Bounds}
\label{sec:conicRadiusCentre}

The matrix inequality~\eqref{eq:gsConicLMI} is nonlinear in $a$ and $b$, and thus solving for an $a$ and $b$ directly given the polytopic plant $\mbc{G}$ is not possible. Three methods of determining tight bounds $a$ and $b$ exists. They consist of
\begin{enumerate}
    \item fixing $a = - \infty$, minimizing $b$, then fixing $b$, and maximizing $a$,
    \item fixing $b = \infty$, maximizing $a$, then fixing $a$, and minimizing $b$, and
    \item defining the conic radius $r = \frac{b-a}{2}$ and conic centre, $c= \frac{a+b}{2}$, and then minimizing the conic radius. 
\end{enumerate}
As~\cite{Bridgeman2014} shows for the LTI case, each method yields different conic bounds. In the polytopic case, the same holds true. Fixing $b = \infty$, and then maximizing $a$ is preferred to determine plant conic bounds because this results in a controller with the least conservative gain. To understand why, note that the controller conic bound $b_{\textrm{c}} = -\frac{1}{a}$ is related to controller gain, and thus an $a$ value as close to zero as possible is desired. When, $b$ is set to $\infty$, Corollary~\ref{col:modifiedConicPoly} is used.

To determine conic bounds using the radius and centre, $r$ and $c$ respectively, consider that 
\beq
    \bma{cc} \mbf{P} \mbf{A}_{i} + \mbf{A}_{i}^\trans \mbf{P} + \mbf{C}_{i}^\trans\mbf{C}_{i}  & \mbf{P} \mbf{B}_{i}- c\mbf{C}_{i}^\trans \\ \star & -\kappa\mbf{1} \ema \leq 0
    \label{eq:polyLMIradius}
\eeq
is equivalent to~\eqref{eq:gsConicLMI}, where $\kappa = -ab$. Note that
\beq
    r^2 = \frac{(a-b)^2}{4} = \frac{(a+b)^2 - 4ab}{4} = c^2 + \kappa,
        \label{eq:radiusObj}
\eeq
and thus minimizing~\eqref{eq:radiusObj} results in the minimum conic radius. The equivalent method for an LTI system is found in~\cite{Joshi2002}. Since~\eqref{eq:radiusObj} is quadratic, it can be transformed to a linear objective along with an LMI constraint by introducing a variable $z$, and then minimizing $z$ subject to the constraint
\beq
    \bma{cc} z - \kappa & c \\ \star & 1 \ema \geq 0,
        \label{eq:zLMISchur}
\eeq
where~\eqref{eq:zLMISchur} is derived from $z \geq c^2 + \kappa$ using the Schur complement~\cite{Boyd:1994kq}.

\section{Design of Polytopic Conic Controllers}
\label{sec:designConicCtrl}


Passivity-based synthesis for affine systems is discussed in~\cite{Walsh2017vsp}. 
The conic-polytopic controller synthesis considered in this paper is an adaptation of the synthesis methods presented in~\cite{Walsh2017vsp,bridgeman2014a,Bridgeman2014}. The goal is to determine a set of controller matrices,
\beq
    \bma{cc} \mbf{A}_{\textrm{c}}(\mbf{s}) & \mbf{B}_{\textrm{c}}(\mbf{s}) \\
        \mbf{C}_{\textrm{c}}(\mbf{s}) & \mbf{0} \ema =
        \sum_{i=1}^N s_i (\mbs{\sigma},\mbf{x},t)
    \bma{cc} \mbf{A}_{\textrm{c},i} & \mbf{B}_{\textrm{c},i} \\
        \mbf{C}_{\textrm{c},i} & \mbf{0} \ema,
        \label{eq:conicCtrlPoly}
\eeq
where the signals $s_i$ satisfy~\eqref{eq:ssAsumptions}. In particular, an $\mc{H}_{\infty}$ controller is synthesized at each vertex of the polytope, and then the closest controller in an $\mc{H}_2$ sense that satisfies conic constraints given by~\eqref{eq:gsConicLMI} is determined.
Associated assumptions are
\begin{packed_enum}
  \item $(\mbf{A}_{i},\mbf{B}_{2,i},\mbf{C}_{2,i})$ is stabilizable and detectable, and
  \item $\mbf{D}_{22} = \mbf{0}$.
\end{packed_enum}
The resulting state space matrices from the $\mc{H}_{\infty}$ synthesis are
$
    (\mbf{A}_{\textrm{c},i} ,\mbf{L}_{i} ,\mbf{K}_{i} , \mbf{0}),
$
where $\mbf{A}_{\textrm{c},i}$ from the $\mc{H}_\infty$ synthesis is also the $\mbf{A}_{\textrm{c},i}$ of the conic controller given in~\eqref{eq:conicCtrlPoly}.
The matrix $\mbf{C}_{\textrm{c},i}$ in~\eqref{eq:conicCtrlPoly} is set to $\mbf{C}_{\textrm{c},i} = \mbf{K}_i$.
The $\mbf{B}_{\textrm{c},i}$ matrix is determined by minimizing the $\mc{H}_2$ norm of the difference between the $\mc{H}_\infty$ controller and a controller that satisfies~\eqref{eq:gsConicLMI}. 
Specifically, this optimization problem is given by minimizing
\beq
  \mc{J}(\mbs{\Pi}_{\textrm{c}},\mbf{B}_{\textrm{c},1},\dots,\mbf{B}_{\textrm{c},N}) = \sum_{i=1}^N\trace (\mbf{B}_{\textrm{c},i} - \mbf{L}_i)^\trans\mbf{W}_i(\mbf{B}_{\textrm{c},i} - \mbf{L}_i),
  \label{eq:conicMinH2B}
\eeq
subject to
\beq
	\bma{ccc}
		\mbs{\Pi}_{\textrm{c}} \mbf{A}_{\textrm{c},i}^\trans 
			+ \mbf{A}_{\textrm{c},i} \mbs{\Pi}_{\textrm{c}}
			&
			\mbf{B}_{\textrm{c},i}  & \mbf{C}_{\textrm{c},i} \mbs{\Pi}_{\textrm{c}} \\
			\star & - \frac{(a_{\textrm{c}} - b_{\textrm{c}})^2}{4 b_\textrm{c}} \mbf{1} &
			-\frac{a_{\textrm{c}} + b_{\textrm{c}}}{2} \mbf{1}\\
			\star & \star & -b_{\textrm{c}} \mbf{1} 	
	\ema < 0,
	\label{eq:conicDesignB}
\eeq
for $i = 1,\dots, N$, where~\eqref{eq:conicDesignB} is derived from Corollary~\ref{col:modifiedConicPoly}. The derivation is omitted for brevity, but see~\cite{Bridgeman2014} for a similar derivation. Equation~\eqref{eq:gsConicLMIMod} cannot be used for controller synthesis since it is bilinear in $\mbf{B}$ and $\mbf{P}$.
The matrix $\mbf{W}_i$ is the observability Grammian that satisfies
$
  \mbf{A}_{\textrm{c},i}^\trans\mbf{W}_i +\mbf{W}_i\mbf{A}_{\textrm{c},i} + \mbf{C}_{\textrm{c},i}^\trans\mbf{C}_{\textrm{c},i} = \mbf{0}.
$
The objective function~\eqref{eq:conicMinH2B} is chosen because when the difference between $\mbf{B}_{\textrm{c}_i}$ and $\mbf{L}_i$ subject to the weight $\mbf{W}_i$ is minimized, the difference between the $\mc{H}_\infty$ controller and the controller that satisfies~\eqref{eq:gsConicLMI} is minimized in an $\mathcal{H}_2$ sense.

The quadratic function~\eqref{eq:conicMinH2B} can be transformed to a linear function and an LMI constraint given by
\beq
	\hat{\mc{J}}(\nu,\mbf{Z},\mbs{\Pi}_{\textrm{c}},\mbf{B}_{\textrm{c},1},\dots,\mbf{B}_{\textrm{c},N}) = \nu,
		\label{eq:objNu}
\eeq
constrained by
\beq
	\nu \geq \trace(\mbf{Z}),
		\label{eq:nuZ}
\eeq
and
\beq
	\bma{cccc} 
		\mbf{Z} & (\mbf{B}_{\textrm{c},1} - \mbf{L}_1)^\trans & \cdots 
			& (\mbf{B}_{\textrm{c},N} - \mbf{L}_N)^\trans \\
		\star & \mbf{W}_1^{-1} & \mbf{0} & \mbf{0} \\
		\vdots & \star & \ddots & \vdots \\
		\star & \star & \star & \mbf{W}_N^{-1}
	\ema \geq 0.
	\label{eq:slackZConstr}
\eeq
Determining $\mbf{B}_{\textrm{c},i}$ is summarized by Problem~\ref{pr:getBci}.

\problem 
\label{pr:getBci}
The problem to determine $\mbf{B}_{\textrm{c},i}$ for $i=1,\dots,N$ is given by minimizing~\eqref{eq:objNu}, subject to~\eqref{eq:conicDesignB},~\eqref{eq:nuZ}, and~\eqref{eq:slackZConstr}. 
\hfill $\Box$

\section{Numerical Example}
\label{sec:numericalExample}

A numerical example using a heat exchanger is used to highlight the usage of the polytopic conic controller. The conic controller is compared to the polytopic LPV controller from~\cite{Apkarian:1995fk}, which uses the Small Gain Theorem for robust closed-loop stability.

\begin{table}[t]
\caption{Heat exchanger properties}
\label{tbl:heatExProp}
\renewcommand{\arraystretch}{1.1}
\centering
\begin{tabular*}{1\columnwidth}{@{\extracolsep{\fill} }l l c c  }
\toprule
Parameter & Unit & Cold Fluid & Hot Fluid \\
\midrule
    $U$ & $\frac{\textrm{J}}{\textrm{s}\cdot \textrm{m}^2 \cdot ^\circ \textrm{C}}$
        & $2411.8$ & $2411.8$ \\ 
    $A$ & $\textrm{m}^2$ & $48.4$ & $48.4$ \\
    $v_{c,\textrm{i}}$, $v_{h,\textrm{i}}$ & $\textrm{m}^3/\textrm{s}$ & $0.04$ & $0.10$ \\
    $v_{c,\textrm{f}}$, $v_{h,\textrm{f}}$ & $\textrm{m}^3/\textrm{s}$ & $0.02$ & $0.06$ \\
    $\rho_c$, $\rho_h$ & $\textrm{kg}/\textrm{m}^3 \cdot 10^3$ & $ 3.50$ & $3.72$ \\
    $c_{pc}$, $c_{ph}$ & $\frac{\textrm{J}}{\textrm{kg} \cdot ^\circ \textrm{C}}$ & $481.8$ & $499.0$ \\
    $V_c$, $V_h$ & $\textrm{m}^3\cdot 10^{-2}$ & $15.8$ & $57.8$ \\
    $T^o_{c,\textrm{i}}$ & $^\circ \textrm{C}$ & $9.3$ & -- \\
    $T^o_{c,\textrm{f}}$ & $^\circ \textrm{C}$ & $25$ & -- \\
\bottomrule
\end{tabular*}
\vspace{-6mm}
\end{table}

\subsection{System Description}

Consider the linearized model of a heat exchanger described by~\cite[pp.~55--60]{Hangos2004}, with properties listed in Table~\ref{tbl:heatExProp}. The nonhomogeneous differential equation describing the dynamics of the heat exchanger is given by
\beq
	\dot{\mbf{T}} = \mbf{A}_{\textrm{p}} \mbf{T} + \mbf{B}_{\textrm{p}} T^i_h + \mbf{W} T^i_c,
		\label{eq:heatExchanger}
\eeq
where
\begin{gather}
	\mbf{A}_{\textrm{p}} = \bma{cc}
			-\frac{v_c(t)}{V_c} - \frac{UA}{c_{pc} \rho_c V_c} &
				 \frac{UA}{c_{pc} \rho_c V_c} 
			\\
				\frac{UA}{c_{ph} \rho_h V_h} &
				-\frac{v_h(t)}{V_h} - \frac{UA}{c_{ph} \rho_h V_h}
		\ema,
        \label{eq:AtimeVarying}
	\\
	\mbf{B}_{\textrm{p}} = \bma{c}
			0
			\\
				\frac{-v_h(t)}{V_h} 
		\ema, \;
	\mbf{W} = \bma{c}
			\frac{-v_c(t)}{V_c} 
			\\
				0
		\ema, \;
	\mbf{T} = \bma{c} T^o_c (t) \\ T^o_h (t) \ema.
    \label{eq:BtimeVarying}
\end{gather}
In this model, the input cold stream $T^i_c$ is constant, and the outlet temperature $T^o_c$ is to be regulated. The hot inlet temperature $T^i_h$ is the control input, and the flow rates of the hot and cold fluids, $v_c(t)$ and $v_h(t)$ respectively, are time varying. In particular, cold and hot stream flow rates are $v_c(t) = \phi(t,v_{c,\textrm{i}},v_{c,\textrm{f}})$ and $v_h(t) = \phi(t,v_{h,\textrm{i}},v_{h,\textrm{f}})$, where $t_{\textrm{f}} = 20$~s, and where
\beq
\begin{split}
    &\phi(t,x_{\textrm{i}},x_{\textrm{f}}) = \\
        &\left\{
            \begin{array}{l l}
                x_{\textrm{i}}, & t \leq 0, \\
                x_{\textrm{i}} + (x_{\textrm{f}} - x_{\textrm{i}})
                    \left[ 3 \left( \frac{t}{t_{\textrm{f}}} \right)^2
                        -2 \left( \frac{t}{t_{\textrm{f}}} \right)^3  \right],
                        & 0 \leq t \leq t_{\textrm{f}}, \\
                x_{\textrm{f}}, & t_{\textrm{f}} \leq t.
            \end{array}
        \right.
\end{split}
\label{eq:phiDef}
\eeq
The desired cold output temperature is given by $T^o_{c,\textrm{d}}(t) = \phi(t,T^o_{c,\textrm{i}},T^o_{c,\textrm{f}})$
To realize a linear system with desired output $y = 0$, the system~\eqref{eq:heatExchanger} is approximated by using the change of variables
$
    \mbf{x} = \mbf{T} + \mbs{\eta} + [ \nu \;\; \nu ]^\trans$, 
    $u = T^i_h + \nu$, and
    $y = \bma{cc} 1 & 0 \ema (\mbf{T} + \mbs{\eta} ) + \nu,
$
where
$
    \mbs{\eta} = (\mbf{A}_{\textrm{p}} + \mbf{W} \bma{cc} 1 & 0 \ema)^{-1} \mbf{W} (T^i_c - T^o_{c,\textrm{f}})$, and
    $\nu = \bma{cc} 1 & 0 \ema \mbs{\eta} - T^o_{c,\textrm{f}}.
$
When fluid-flow is constant, $\dot{\mbf{x}} = \dot{\mbf{T}}$. The resulting state-space realization is determined using~\eqref{eq:phiDef}. The matrices $\mbf{A}$ and $\mbf{B}_2$ are
\bdis
    \mbf{A} = \sum_{i=1}^2 s_i(\mbs{\sigma},\mbf{x},t) \mbf{A}_i, \quad \mbf{B}_2 = \sum_{i=1}^2 s_i(\mbs{\sigma},\mbf{x},t) \mbf{B}_{2,i},
\edis
where $s_1 = \phi(t,1,0)$, and $s_2 = 1 - s_1 = \phi(t,0,1) $. The matrices $\mbf{A}_1$ and $\mbf{B}_{2,1}$ are defined by evaluating $\mbf{A}_{\textrm{p}}$ and $\mbf{B}_{\textrm{p}}$ in~\eqref{eq:AtimeVarying}-\eqref{eq:BtimeVarying} at $v_{c,\textrm{i}}$ and $v_{h,\textrm{i}}$. The matrices $\mbf{A}_2$ and $\mbf{B}_{2,2}$ are defined by evaluating $\mbf{A}_{\textrm{p}}$ and $\mbf{B}_{\textrm{p}}$ in~\eqref{eq:AtimeVarying}-\eqref{eq:BtimeVarying} at $v_{c,\textrm{f}}$ and $v_{h,\textrm{f}}$. The matrix $\mbf{C}_2$ is constant, where $ \mbf{C}_2 = [1 \;\; 0 ]$.

\subsection{Robust Control Design}

The parameters in $\mbf{A}_{\textrm{p}}$ are generally uncertain. 
For example, $U$ is uncertain due to CaCO$_3$ formation \cite[p.~31]{Hangos2004}. The uncertain $\mbfbar{A}$ matrix is modelled as
$
	\mbfbar{A} = \mbf{A} + \mbf{A}_\delta,
$
where
\bdis
	\mbf{A}_\delta = \delta \bma{cc}
			\frac{UA}{c_{pc} \rho_c V_c} &
				-\frac{UA}{c_{pc} \rho_c V_c} 
			\\
				-\frac{UA}{c_{ph} \rho_h V_h} &
				\frac{UA}{c_{ph} \rho_h V_h}
		\ema,  \quad \delta \in \mathbb{R}.
\edis
The system for robust control design is then
\begin{align}
	\dot{\mbf{x}} &= \mbf{A} \mbf{x} + \mbf{B}_2 \mbf{u} 
		+ \mbf{q}, \quad
	\mbf{y} = \mbf{C}_2 \mbf{x}, \quad
	\mbf{p} =  \mbf{x} ,
		\label{eq:robSysDesign3} 
\end{align}
where the uncertainty block is given by $\mbf{q} = \mbf{A}_\delta \mbf{p}$, and thus $\mbs{\Delta} = \mbf{A}_\delta$. The weighting matrices for robust control design using~\eqref{eq:robSysDesign3} are
\beq
    \mbf{B}_3 = \mbf{1}, \quad \mbf{C}_3 = \mbf{1}, \quad \mbf{D}_{23} = \mbf{0},
        \quad \mbf{D}_{32} = \mbf{0}.
        \label{eq:robustWeights}
\eeq
Other weighting combinations are possible, such as by factoring $\mbf{A}_\delta$, with factorization  $\mbf{B}_3 \mbf{C}_3 = \mbf{A}_\delta$ or $\mbf{B}_3 \mbf{C}_3 = \delta^{-1}\mbf{A}_\delta$, and then by setting $\mbs{\Delta} = \mbf{1}$ or $\mbs{\Delta} = \delta\mbf{1}$. In some cases, this may provide a numerically simpler controller synthesis, but in the heat exchanger example, no significant controller improvements were found.
Table~\ref{tbl:heatExCones} shows the heat exchanger's conic sectors and $||\mbs{\Delta}||_\infty$ for various values of $\delta$. A value of $\delta = 0.5$ is equivalent to halving the heat exchanger's area, such as during CaCO$_3$ buildup. A value of $\delta = -1$ doubles the heat exchanger's area.

\begin{table}[t!]
\caption{Conic bounds of the heat exchanger model.}
\label{tbl:heatExCones}
\renewcommand{\arraystretch}{1.1}
\centering
\begin{tabular*}{1\columnwidth}{@{\extracolsep{\fill} }l c c c c }
\toprule
Plant & Conic Max $a$ & Conic Min $r$  &  $\norm{\mbs{\Delta}}_\infty$ \\
\midrule
    $\delta = 0$ & $\cone[   -0.06,   98.9]$    & 
        $\cone[   -0.14 ,   0.38]$ & $0$ \\ 
    $\delta = 0.5$ & $\cone[     -0.04,    97.4]$ & 
        $\cone[   -0.09 ,   0.24]$ & $0.057$ \\
    $ \delta = -1$ & $\cone[   -0.08 ,   99.4]$ & 
        $\cone[   -0.19 ,   0.52]$ & $0.23$ \\
\bottomrule
\end{tabular*}
\end{table}

Using the design method for an LPV controller from~\cite{Apkarian:1995fk}, the input must be filtered to obtain a constant $\mbf{B}_2$ matrix. A first-order filter with a cutoff frequency of $2$~rad/s is implemented on the input of the system, which adds system states. Designing the polytopic LPV controller using the weights from~\eqref{eq:robustWeights} results $|| \mbc{G}_{\textrm{cl}} ||_{\infty} = 16.67$. Recall that for robust closed-loop stability via the Small Gain Theorem, $|| \mbc{G}_{\textrm{cl}} ||_{\infty} < 1/||\mbs{\Delta}||_\infty$. This means that stability can be guaranteed via the Small Gain Theorem for $\delta = 0.5$, but cannot be guaranteed for $\delta = -1$.

\begin{figure}[t!]
    \centering
    \includegraphics[width=\columnwidth]{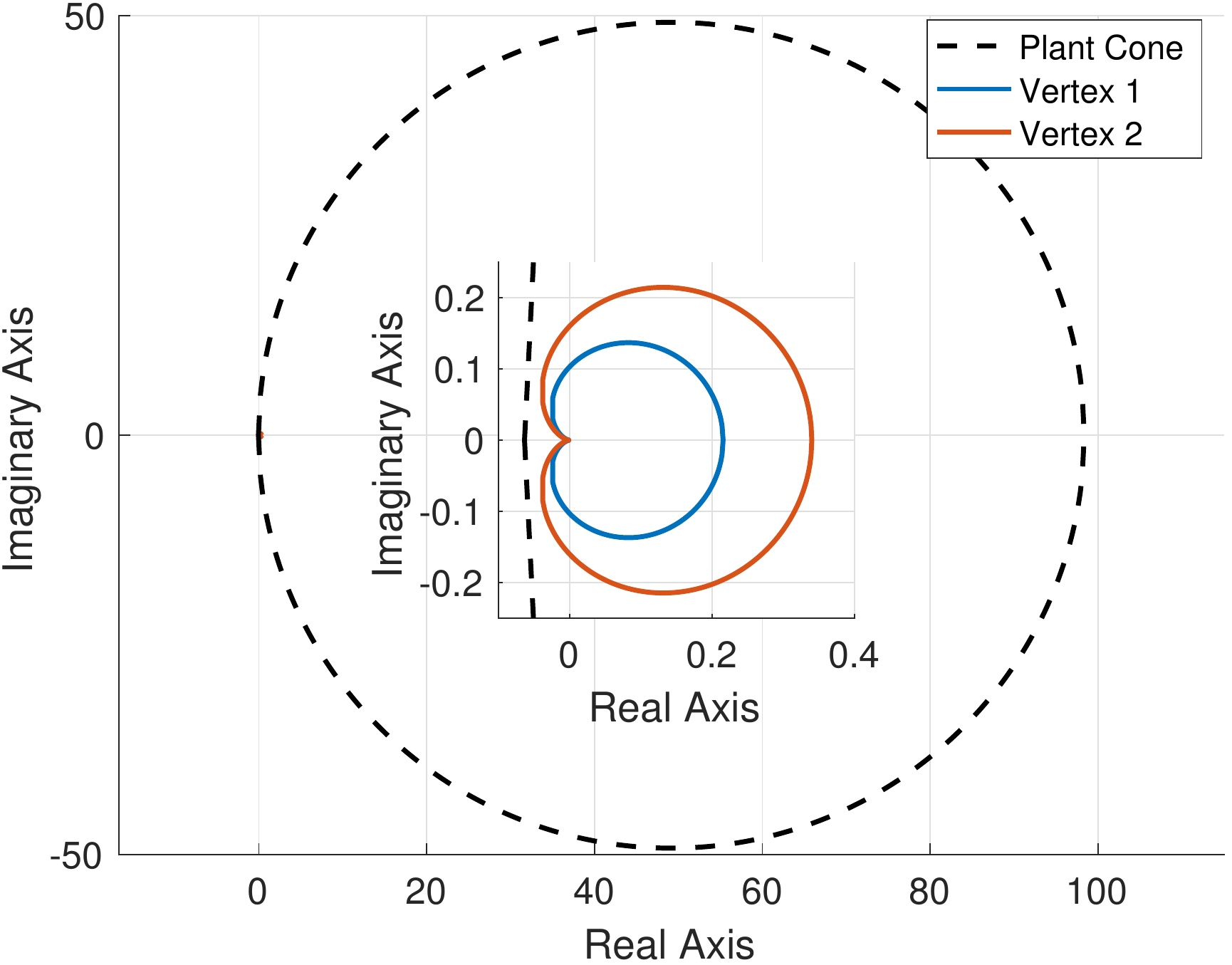}
    \caption{Plant conic boundary and plant Nyquist plot. Embedded plot is a zoomed-in version to emphasize the plant's Nyquist plot at the vertices.}
\label{fig:plantConic}
\vspace{-3mm}
\end{figure}

\begin{figure}[t!]
    \centering
    \includegraphics[width=\columnwidth]{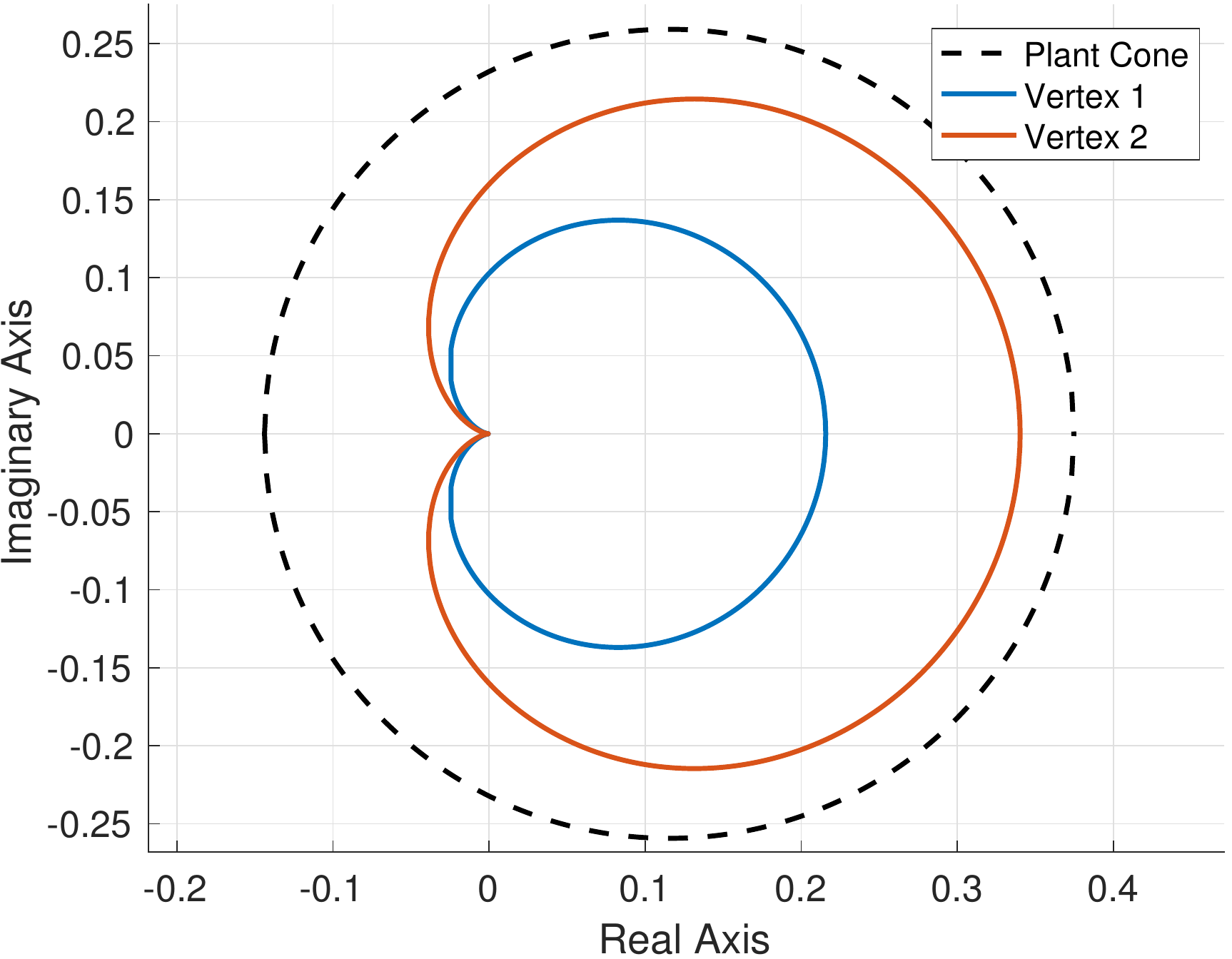}
    \caption{Plant conic boundary and plant Nyquist plot with minimum conic radius.}
    \label{fig:plantMinR}
    \vspace{-3mm}
\end{figure}

The conic bounds of the heat exchanger are found using Theorem~\ref{thm:conicPolytopic}.
The Nyquist plots of $\mbc{G}$ at the vertices of the polytope are shown in Fig.~\ref{fig:plantConic}, as well as a circle that denotes the conic bounds. Notice that the plant lies well within the conic bounds determined by Theorem~\ref{thm:conicPolytopic}. The Nyquist plot of the vertices is shown in Fig.~\ref{fig:plantMinR} when the conic radius is minimized. In this plot, the black dashed circle is much tighter than when maximizing $a$. This would seem like tighter conic bounds and thus improved controller performance, but the more negative $a$ leads to a decreased $b_{\textrm{c}}$, and thus yields a controller with more conservative gain and inferior performance. 
To robustly stabilize the plant, the largest cone that captures both the nominal and perturbed plant must be considered, thus the cone for controller design is $\cone[-0.19,0.52]$ for minimizing $r$ and $\cone[-0.08,99.4]$ for maximizing $a$.

The weights given by~\eqref{eq:robustWeights} are used to design the $\mc{H}_\infty$ controllers at each vertex. These controllers are used for controller synthesis, as discussed in Section~\ref{sec:designConicCtrl}.
To analyze the utility of forcing the $\mc{H}_\infty$ controllers at each vertex to satisfy Theorem~\ref{thm:concSector}, the $\mc{H}_\infty$ controllers at each vertex are linearly interpolated to form an $\mc{H}_\infty$ gain-scheduled controller. This controller has no stability or performance guarantees when in closed-loop with the heat exchanger. 

\begin{figure}[t!]
    \centering
    \includegraphics[width=\columnwidth]{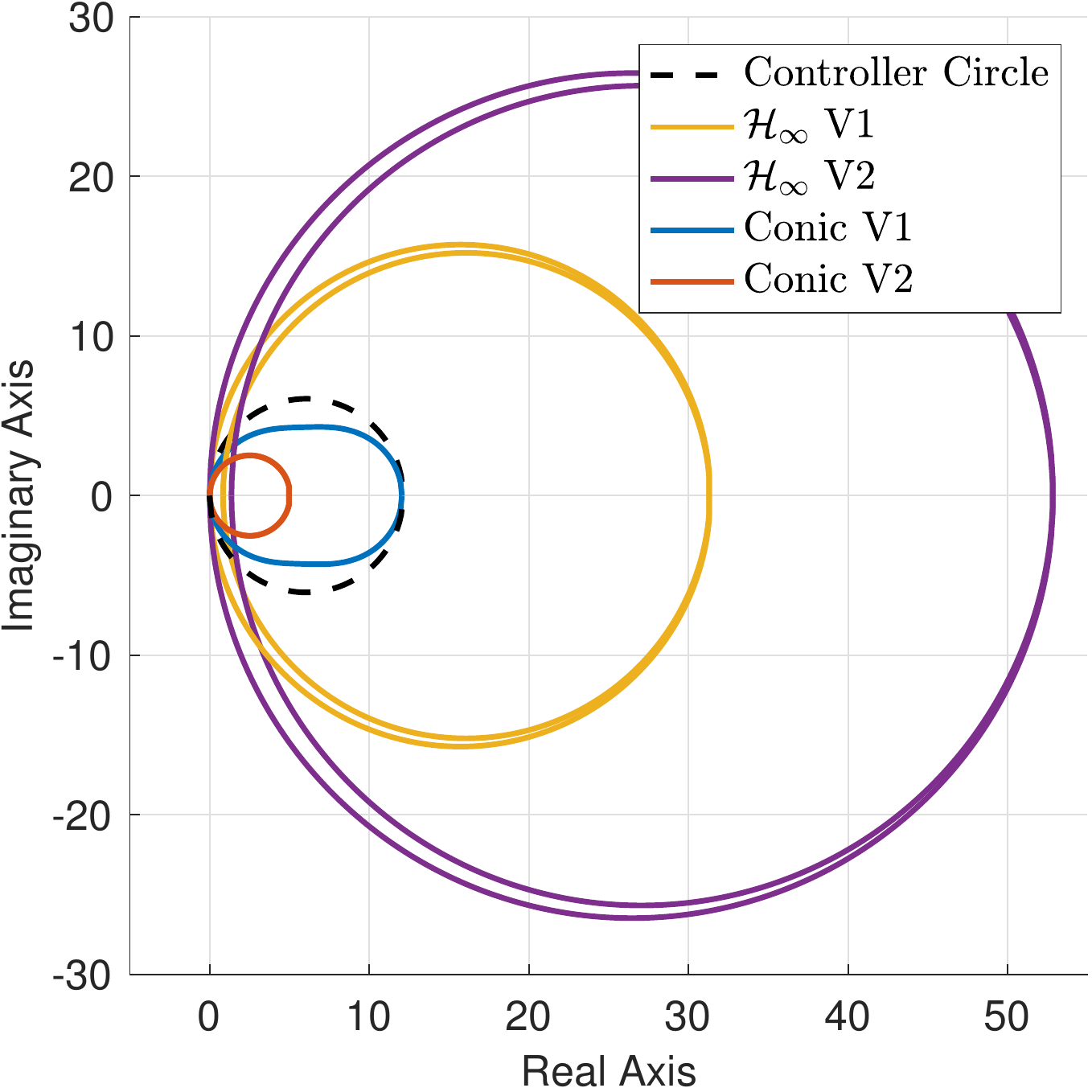}
\caption{Nyquist plot of $\mc{H}_\infty$ controller and the conic controller at each vertex of the polytope. The controller circle plots the conic bounds of the controller.}
\label{fig:ctrlNyquist}
\end{figure}

The Nyquist plots of the $\mc{H}_\infty$ controller at each vertex and the conic controller at each vertex, synthesized using maximum plant $a$, are shown in Fig.~\ref{fig:ctrlNyquist}. The gain of the interpolated $\mc{H}_\infty$ controller is much larger than the gain of the conic controllers at each vertex. When synthesizing the conic controller, the effect of the synthesis decreases gain of the controller so that the controller lies within $\cone(a_{\textrm{c}},b_{\textrm{c}})$. The gain at both vertices are not the same. The gain at vertex 1 of the $\mc{H}_{\infty}$ controller is smaller than at vertex 2, and this is reversed for the conic case.

\begin{figure}[t!]
    \centering
    \includegraphics[width=\columnwidth]{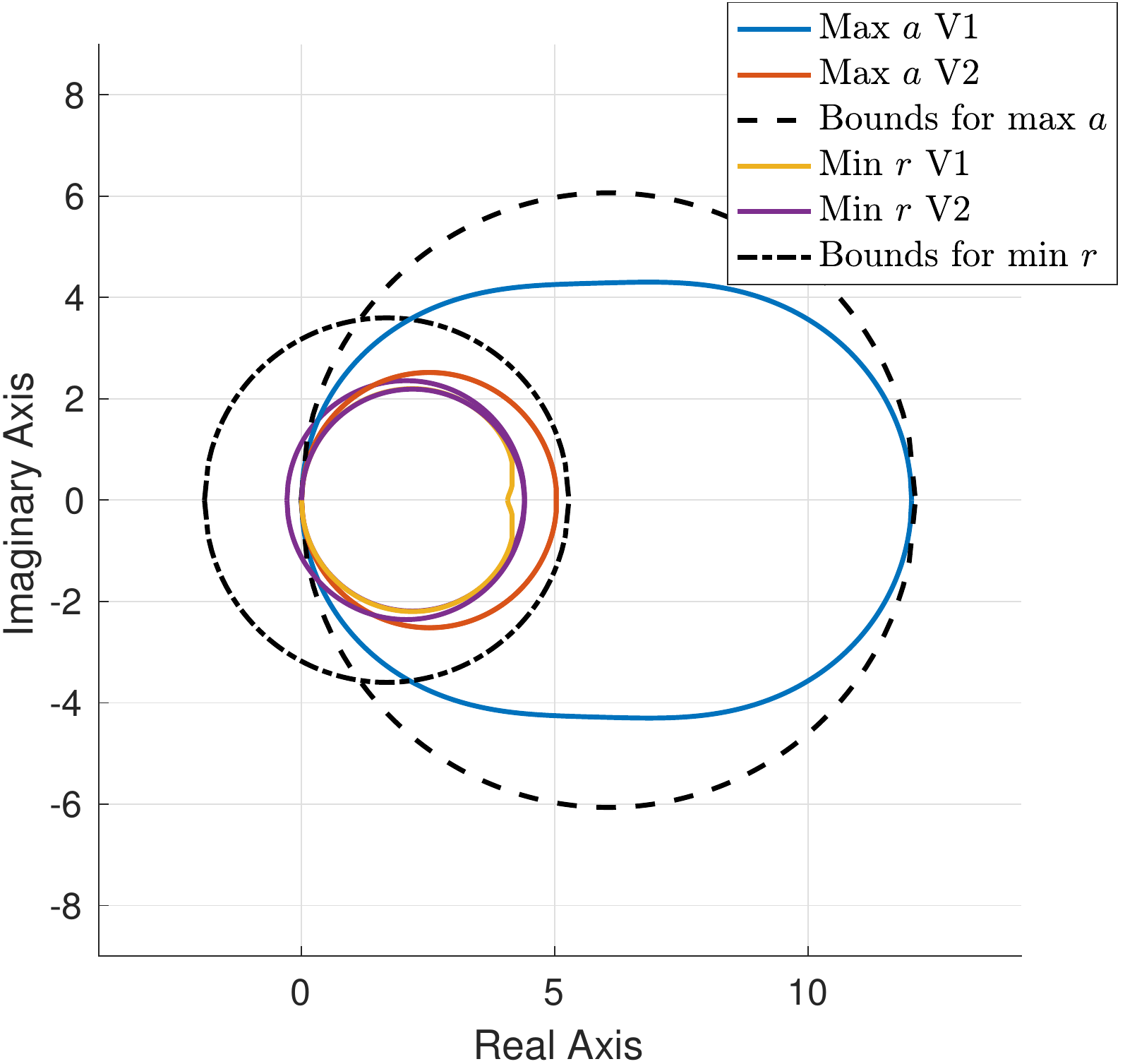}
    \caption{Nyquist plot of controllers at vertices of polytope synthesized using maximum plant $a$ and minimum plant $r$ conic bounds.}
    \label{fig:ctrlMaxAMinR}
    \vspace{-5mm}
\end{figure}

Fig.~\ref{fig:ctrlMaxAMinR} shows the conic bounds for the controller synthesized with maximum plant $a$ and minimum plant $r$. The $b_{\textrm{c}}$ bound, which is related to the upper limit on controller gain, is approximately $2.5$ times larger when using the bounds determined by maximum $a$ than by minimum $r$. It is expected that the controller with the larger design freedom exhibits superior closed-loop performance, that is the RMS error of $T^o_c$ should be less for the max $a$ controller than the min $r$ controller.


\subsection{Numerical Results}
\label{sec:numResultsHeat}

\begin{figure*}[t]
\centering
\begin{subfigure}{0.49\textwidth}
    \centering
    \includegraphics[width=.95\columnwidth]{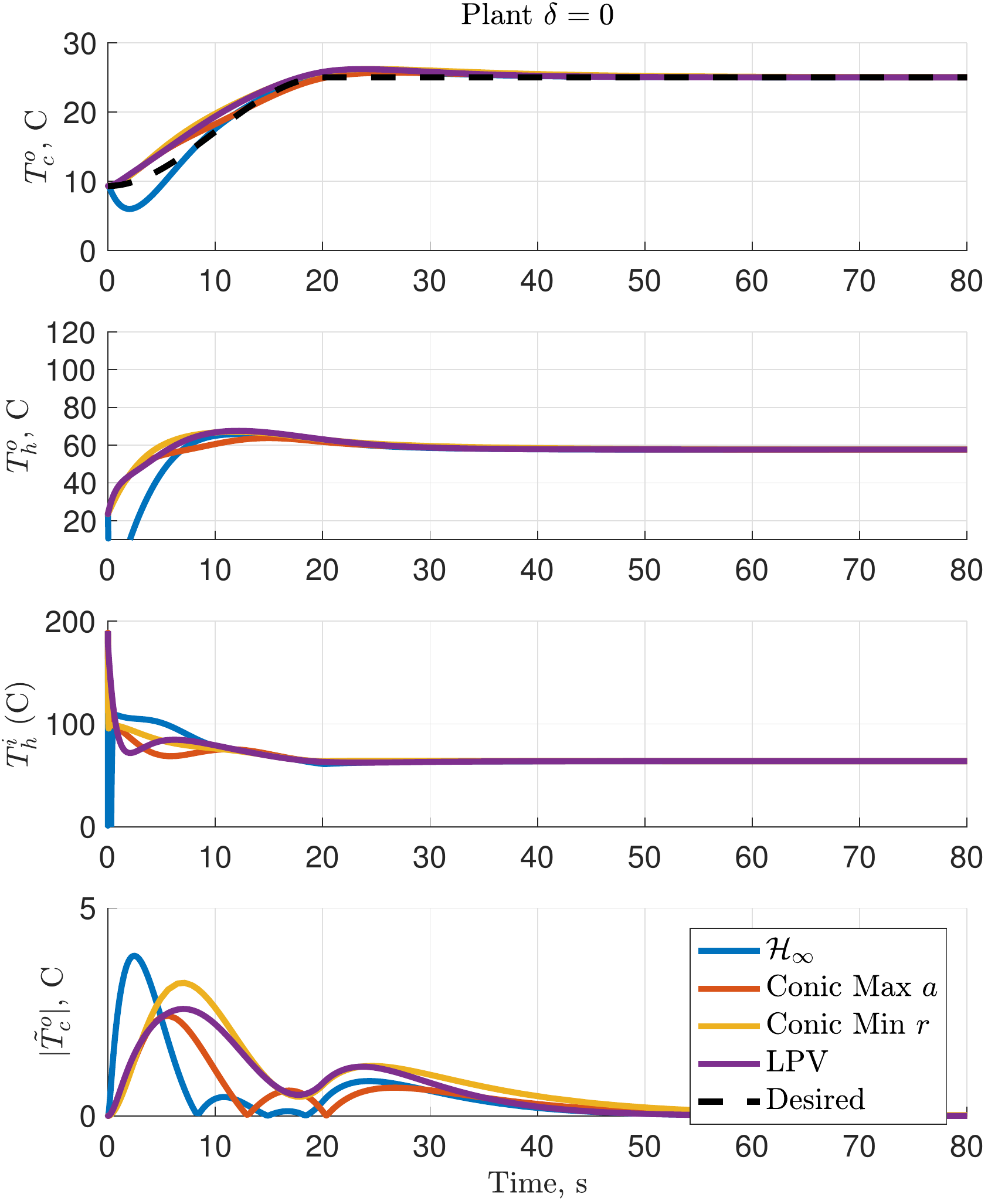}
    \caption{Nominal plant with $\delta = 0$.}
    \label{fig:nomPlant}
\end{subfigure}
\begin{subfigure}{0.49\textwidth}
    \centering
    \includegraphics[width=.9595\columnwidth]{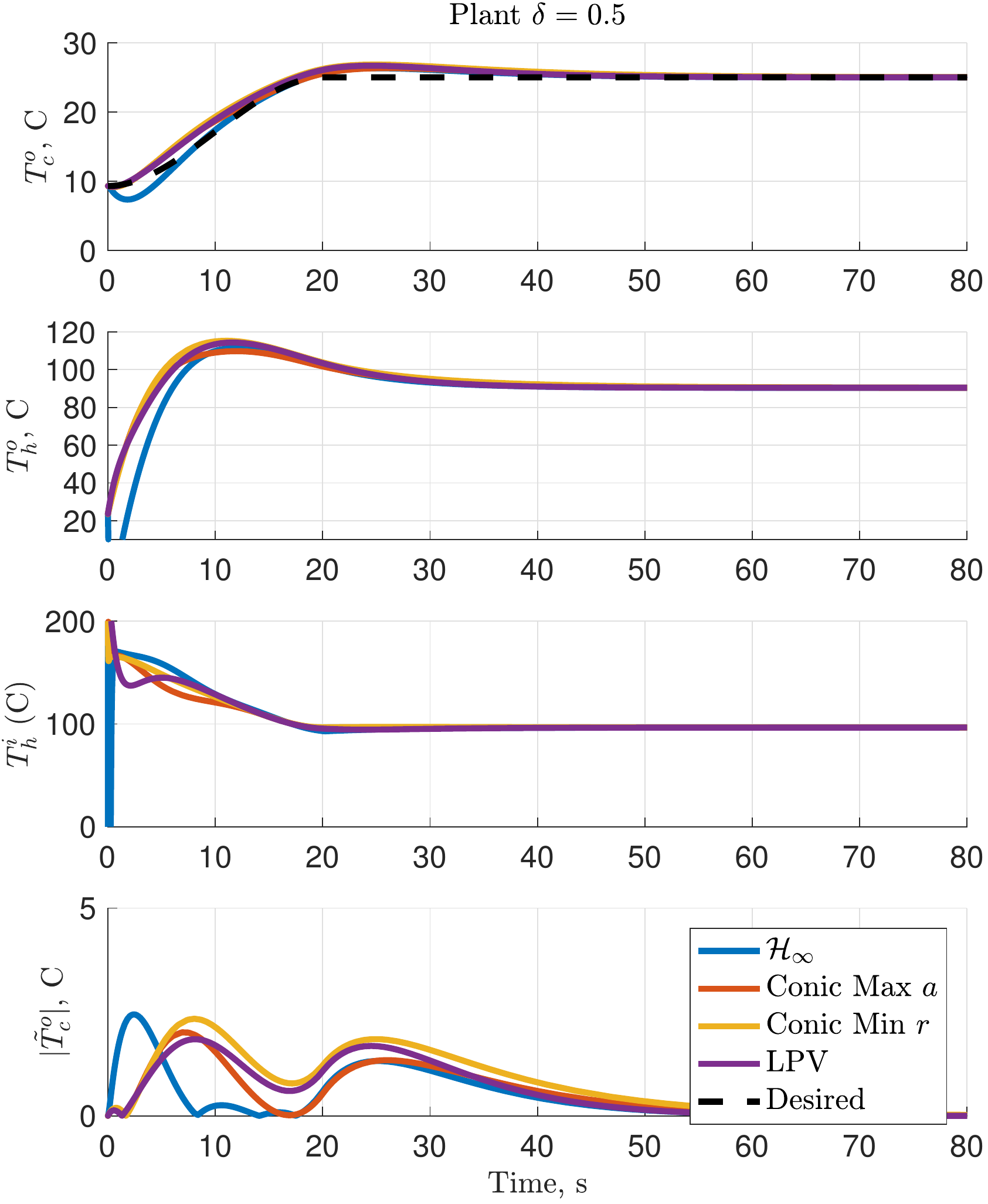}
    \caption{Modified plant with $\delta = 0.5$.}
    \label{fig:modPlant}
\end{subfigure}
\caption{Simulation results to track $T^o_c$, with input $T^i_h$ of the heat exchanger.}
\label{fig:simResults}
\vspace{-5mm}
\end{figure*}

Numerical results are shown in Fig.~\ref{fig:simResults}, and the case $\delta = -1$ is omitted for brevity. The root-mean square (RMS) values of the error of the cold outlet temperature are shown in Table~\ref{tbl:heatExErr}. At $\delta = 0.5$, the area for heat exchange halves. 
The performance of all controllers improve with $\delta = 0.5$, which implies that not all uncertainty degrades performance. When $\delta = -1$, the opposite is true and the performance of all controllers deteriorates. However, comparing the change of performance of each controller with respect to $\delta$ yields interesting results. The standard deviation between the RMS errors is seven times less for the conic controller design than with the $\mc{H}_\infty$-based LPV design. This is partly expected because conic controllers exhibit a similar level of robustness for linear control design~\cite{Bridgeman2014}. This lack of sensitivity to model uncertainty highlights the benefits of conic-sector-based control. 

The conic controller that uses the maximum $a$ plant sector has superior performance that the conic control that uses the minimum $r$ plant sector. This is expected since the maximum $a$ plant sector controller has larger conic bounds, and thus less design restriction. Both conic controllers have superior performance compared to the LPV controller, but this can be explained since the LPV controller was designed to maximize robustness instead of performance.

For $\delta = 0$ and $\delta = 0.5$, the $\mc{H}_\infty$ controller actually performs better than the LPV controller. This may be explained by the fact that for polytopic controller design, a common Lyapunov matrix is required, which is a source of conservatism. The interpolated $\mc{H}_\infty$ is synthesized at the vertices, and even though it has no guarantee to stabilize the closed-loop system, instability was not the result. However, with a different set of weighting matrices, in simulation, the interpolated $\mc{H}_\infty$ controller lead to poor closed-loop performance, and even closed-loop instability.

\begin{table}[t!]
\caption{RMS error of $T^o_c - T^o_{c,\textrm{d}}$, $^\circ C$}
\label{tbl:heatExErr}
\renewcommand{\arraystretch}{1.1}
\centering
\begin{tabular*}{1\columnwidth}{@{\extracolsep{\fill} }l c c c c }
\toprule
Plant & $\mc{H}_\infty$ & Conic Max $a$ & Conic Min $r$  &  LPV \\
\midrule
    $\delta = 0$ & $1.13$    & $0.612$ & $0.912$ & $1.19$ \\ 
    $\delta = 0.5$ & $0.782$ & $0.606$ & $0.850$ & $0.832$ \\
    $\delta = -1$ & $1.72$ & $0.721$ & $1.047$ & $1.68$ \\
\midrule
    std. dev. & $0.477$ & $0.065$ & $0.100$ & $0.424$ \\
\bottomrule
\end{tabular*}
\end{table}

\section{Concluding Remarks}
\label{sec:conclusion}

This paper provides LMI conditions to determine conic bounds for interior-conic polytopic systems. This paper also provides a method that synthesizes a polytopic controller subject to conic bounds. In simulation, the synthesized conic controllers exhibit very little sensitivity to model uncertainty. 

For conic-sector-based robust control, the controller is forced to satisfy conic bounds to ensure closed-loop stability with $\mbc{G}_\Delta$. For traditional robust control, the controller is designed such that $\mbc{G}_{\textrm{cl}}$ satisfies the small gain condition with respect to the uncertainty block, $\mbs{\Delta}$. Since the conic-sector and small gain methods approach robustness in a different manner, this paper only provides an indirect comparison.
In the future, it is of interest to design $\mbc{G}_{\textrm{cl}}$ to satisfy conic bounds, or to have $\mbc{G}_{\textrm{c}}$ satisfy the small gain condition with respect to $\mbc{G}_\Delta$ so that uncertainty is approached in a similar manner.

A common Lyapunov matrix is required for both these methods, which is a source of conservatism when determining conic bounds for the plant, and when synthesizing the controller. This conservatism may provide undesirably large conic bounds for the plant, which would lead to less control design freedom. When synthesizing the conic controller, the common Lyapunov matrix may prevent the controller to use the full allowable controller cone.
The effects and methods to mitigate the common Lyapunov matrix is an area of future research.

\appendix[Proof of Lemma~\ref{lem:yNorms}]

This Appendix presents the proof of Lemma~\ref{lem:yNorms}.
Each element of $\mbf{y}$ is given by
$
	y_j(t) = \sum_{i=1}^N s_i(\mbs{\sigma},\mbf{x},t) y_{ij}(t),
$
whose square results in
\begin{align*}
	y_j^2(t) &= \abs{\sum_{i=1}^N s_i(\mbs{\sigma},\mbf{x},t) y_{ij}(t)}^2, \\
		&= \abs{\sum_{i=1}^N \sqrt{s_i(\mbs{\sigma},\mbf{x},t)} \left(\sqrt{s_i(\mbs{\sigma},\mbf{x},t)} y_{ij}(t) \right)}^2.
\end{align*}
Using the Cauchy-Schwartz Inequality yields
\begin{align*}
	y_j^2(t) &\leq
			\left(\sum_{i=1}^N \abs{\sqrt{s_i(\mbs{\sigma},\mbf{x},t)}}^2 \right) 
			\left( \sum_{i=1}^N \abs{\sqrt{s_i(\mbs{\sigma},\mbf{x},t)} y_{ij}(t)}^2\right) \\
		&= \left(\sum_{i=1}^N s_i(\mbs{\sigma},\mbf{x},t) \right) 
			\left( \sum_{i=1}^N (\sqrt{s_i(\mbs{\sigma},\mbf{x},t)} y_{ij}(t))^2\right) \\
		&= \left( \sum_{i=1}^N s_i(\mbs{\sigma},\mbf{x},t) y_{ij}^2(t)\right).
\end{align*}
Taking the sum from $j=1$ to $m$, and then rearranging yields
\bdis
	\sum_{j=1}^m y_j^2(t) \leq \sum_{j=1}^m  \left( \sum_{i=1}^N s_i(\mbs{\sigma},\mbf{x},t) y_{ij}^2(t)\right)
\edis
\begin{align*}
	\mbf{y}^\trans(t) \mbf{y}(t) &\leq 
		  \sum_{i=1}^N s_i(\mbs{\sigma},\mbf{x},t)  \sum_{j=1}^m  y_{ij}^2(t) \\
		  &= 
		  \sum_{i=1}^N s_i(\mbs{\sigma},\mbf{x},t)  \mbf{y}_i^\trans(t) \mbf{y}_i(t). 
\end{align*}
Integrating both sides in time from $0$ to $T$ in time yields
\begin{align}
	\int_{0}^T \mbf{y}^\trans(t) \mbf{y}(t) \; \dt &\leq
		\sum_{i=1}^N \int_{0}^T  s_i(\mbs{\sigma},\mbf{x},t)  \mbf{y}_i^\trans(t) \mbf{y}_i(t) \; \dt \notag \\
	\norm{\mbf{y}(t)}_{2T}^2 &\leq \sum_{i=1}^N \norm{\sqrt{s_i(\mbs{\sigma},\mbf{x},t)} \mbf{y}_i(t)}_{2T}^2.
		\label{eq:yNorms}
\end{align}
Multiplying both sides of~\eqref{eq:yNorms} by $-1$ yields~\eqref{eq:yNormsNeed}.

\bibliographystyle{ieeetr}

\bibliography{refConic}  

\end{document}